\documentclass[amsmath,amssymb,aps,prl,nofootinbib,superscriptaddress,twocolumn,preprintnumbers]{revtex4-2}
\usepackage[utf8]{inputenc}
\usepackage{mathrsfs}
\usepackage{bm}
\usepackage{url}
\usepackage[normalem]{ulem}
\usepackage{mathtools}
\usepackage{array}
\newcolumntype{P}[1]{>{\centering\arraybackslash}p{#1}}
\newcolumntype{M}[1]{>{\centering\arraybackslash}m{#1}}
\usepackage[caption=false]{subfig}
\usepackage{dcolumn}
\usepackage{graphicx, epsfig}
\usepackage[dvipsnames]{xcolor}
\usepackage{mathrsfs}
\usepackage{bm}
\usepackage{yhmath}
\usepackage[caption=false]{subfig}
\usepackage[normalem]{ulem}
\usepackage{mathtools}
\usepackage{bigints}
\usepackage{float}
\usepackage[colorlinks = true, linkcolor = purple, urlcolor  = blue, citecolor = blue, anchorcolor = blue]{hyperref}
\usepackage{float}
\usepackage{multirow}
\usepackage{tikz,xcolor,hyperref}
\definecolor{darkgreen}{rgb}{0.0, 0.2, 0.13}
\definecolor{bostonuniversityred}{rgb}{0.8, 0.0, 0.0}
\definecolor{lime}{HTML}{A6CE39}
\DeclareRobustCommand{\orcidicon}{
	\begin{tikzpicture}
	\draw[lime, fill=lime] (0,0) 
	circle [radius=0.16] 
	node[white] {{\fontfamily{qag}\selectfont \tiny ID}};
	\draw[white, fill=white] (-0.0625,0.095) 
	circle [radius=0.007];
	\end{tikzpicture}
	\hspace{-2mm}
}
\foreach \x in {A, ..., Z}{\expandafter\xdef\csname orcid\x\endcsname{\noexpand\href{https://orcid.org/\csname orcidauthor\x\endcsname}
			{\noexpand\orcidicon}}
}
\newcommand{\be}{\begin{equation}}
\newcommand{\ee}{\end{equation}}
\newcommand{\ba}{\begin{eqnarray}}
\newcommand{\ea}{\end{eqnarray}}

\def\lp4{$\lambda \phi^4$}
\begin{document}
\preprint{\texttt{FERMILAB-PUB-26-0621-T}}
\title{Getting Warmer: IceCube Nears Freeze Out}
\author{Mainak Mukhopadhyay\hspace{-1mm}\orcidA{}}
\email{mainak@fnal.gov}
\affiliation{Theoretical Astrophysics Department, Theory Division, Fermi National Accelerator Laboratory, Batavia, Illinois 60510, USA}
\affiliation{Kavli Institute for Cosmological Physics, University of Chicago, Chicago, Illinois 60637, USA}
\affiliation{NSF-Simons AI Institute for the Sky (SkAI), 172 E. Chestnut St., Chicago, Illinois 60611, USA}
\author{Gordan Krnjaic\hspace{-1mm}\orcidB{}}
\email{krnjaicg@fnal.gov}
\affiliation{Theoretical Astrophysics Department, Theory Division, Fermi National Accelerator Laboratory, Batavia, Illinois 60510, USA}
\affiliation{Department of Astronomy and Astrophysics, University of Chicago, Chicago, Illinois 60637, USA}
\affiliation{Kavli Institute for Cosmological Physics, University of Chicago, Chicago, Illinois 60637, USA}
\date{\today}
\begin{abstract}
IceCube has recently detected a diffuse population of high-energy neutrinos arising from the Milky Way. We use this high-significance detection  to place new limits on dark matter (DM) annihilation to neutrinos with two complementary approaches. The first method uses the background-subtracted Galactic longitude distribution of shower events to place a conservative bound on the DM annihilation cross section that does not rely on any assumed Galactic cosmic ray emission model; the resulting limits on the velocity-averaged annihilation cross section improve upon existing bounds by factors of a few. The second method uses the template-dependent neutrino energy spectra from the Inner Galaxy, inferred under different Galactic cosmic ray emission models. 
This complementary approach shows that the inferred Galactic neutrino intensities are already sensitive to DM contributions near the thermal-relic benchmark for a range of TeV-scale DM masses, though this comparison  is more model-dependent. Our results demonstrate that measurements of diffuse Galactic neutrino emission can be used as a powerful probe of DM annihilation into neutrinos. Future observations with IceCube-Gen2 and KM3NeT will substantially extend this sensitivity, potentially allowing a decisive test of the thermal freeze-out mechanism with Galactic neutrino observations.
\end{abstract}
\maketitle
\section{Introduction}
The particle nature of dark matter (DM) remains one of the central open questions in fundamental physics~\cite{Bertone:2004pz}. A particularly compelling possibility is that DM was once in chemical equilibrium with the Standard Model (SM) such that its present-day abundance is set by its annihilation cross section~\cite{Lee:1977ua}. For weak-scale DM, the observed relic density is achieved for a characteristic velocity-averaged annihilation cross section of $\langle \sigma v \rangle_{\rm th} \approx 2 \times 10^{-26}\ {\rm cm^3 s^{-1}}$~\cite{Steigman:2012nb}, providing a well-motivated benchmark for indirect-detection searches. Thermal production therefore motivates searches for the products of ongoing DM annihilation in the present-day Universe. Since the annihilation signal depends quadratically on the DM density, the inner Milky Way, where Galactic DM is most concentrated, offers a powerful target for DM annihilation searches.

Neutrinos provide a particularly important probe of freeze out~\cite{Arguelles:2019ouk}. While annihilation to quarks, charged leptons, or electroweak bosons generally produces electromagnetic signatures, there are many viable models in which  DM couples predominantly to neutrinos~\cite{Blennow:2019fhy}. Such scenarios arise, for example, in models connecting the dark sector to neutrino mass generation~\cite{Boehm:2006mi,Farzan:2012sa}. Consequently, annihilation to neutrino-antineutrino pairs constitutes an especially important benchmark for indirect-detection searches, representing one of the most experimentally elusive Standard Model final states and providing a particularly model-independent probe of DM annihilation~\cite{Beacom:2006tt}. 

Despite extensive searches with neutrino telescopes, probing the canonical thermal relic annihilation cross section to neutrinos has remained elusive.\footnote{Technically, Super Kamiokande has excluded a thermal relic annihilation cross section  for a  very narrow range of DM masses near $\sim$ 30 MeV \cite{Arguelles:2019ouk}.} 
IceCube has searched for neutrino signals from DM annihilation in the Galactic Center and Galactic halo~\cite{IceCube:2011kcp,IceCube:2015rnn,IceCube:2023ies,IceCube:2025fcn}, and the Sun~\cite{IceCube:2021xzo}, while complementary searches have been performed by ANTARES~\cite{ANTARES:2015vis,Gozzini:2023nka} and KM3NeT/ARCA~\cite{KM3NeT:2024xca}. Although these searches have significantly improved sensitivity to DM annihilation into neutrinos over the past decade, the canonical thermal relic benchmark has remained largely beyond reach for TeV-scale DM. 

Recently, IceCube reported the first high-significance detection of diffuse high-energy neutrino emission from the Galactic Plane at a post-trial significance of $5.7\sigma$, with the observed emission strongly concentrated toward the Inner Galaxy~\cite{IceCube:2026plr}. Although this emission is expected to arise predominantly from interactions of Galactic CRs (cosmic rays) with interstellar gas,  any additional neutrino component from DM annihilation must also contribute to the measured Galactic signal. Importantly, the new observations provide both template-dependent estimates of the diffuse Galactic neutrino intensity and the background-subtracted spatial distribution of shower and track events across reconstructed Galactic longitude. Together, these complementary observables enable powerful new tests of DM annihilation in the Inner Galaxy.

\begin{figure}[ht!]
\centering
\hspace{-0.5cm}
\includegraphics[width=0.49\textwidth]{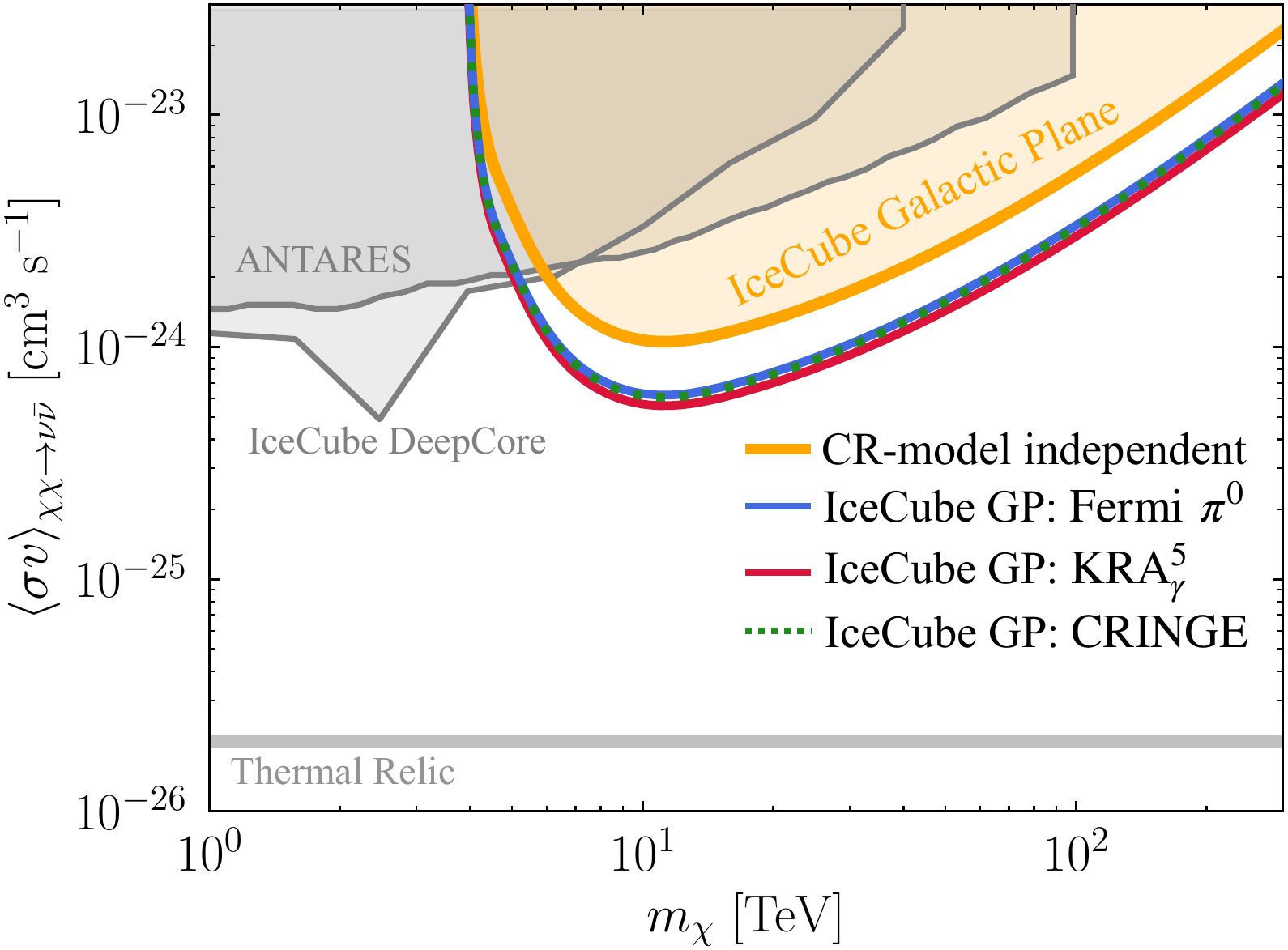}
\caption{\label{fig:ann_lim_chisq} Constraints on the $\chi \chi \rightarrow \nu \bar{\nu}$ annihilation cross section as a function of DM mass $m_\chi$, derived from the IceCube Galactic Plane (GP) observations and assuming a NFW Galactic DM density profile. The model-independent $90\%$ CL limit labeled IceCube GP is derived from the reconstructed Galactic longitude distribution of the background-subtracted shower events using Eqs.~\eqref{eq:N_i} - \eqref{eq:chi2-criterion}. The colored curves show the corresponding model-dependent limits after subtracting the different Galactic CR templates; note that the limits from subtracting the Fermi $\pi^0$ (blue) and CRINGE (green) templates are nearly indistinguishable on this plot. Existing $90\%$ CL limits from IceCube (DeepCore)~\cite{IceCube:2023ies}, ANTARES~\cite{ANTARES:2015vis,Albert:2016emp}, and the benchmark thermal relic annihilation cross section are also shown.
}
\end{figure}

In this \emph{Letter}, we show that the latest high-energy neutrino observations of the Inner Galaxy are approaching sensitivity to the thermal relic cross section for DM annihilating directly to neutrinos. Assuming a Navarro–Frenk–White (NFW) Galactic DM halo, we compare $\chi\chi\rightarrow\nu\bar{\nu}$ annihilation against recent 
IceCube results using two complementary methods:
\begin{itemize}
    \item{\bf Spatial Analysis:} we set new limits on $\langle \sigma v\rangle$ as a function of $m_\chi$ using the background-subtracted reconstructed Galactic longitude distribution of shower events, both before and after subtracting various CR model templates for Galactic neutrino emission.
    
    \item{\bf Energy Spectrum Analysis:} we compare neutrino intensities from DM annihilation against various best-fit spectral templates for Galactic neutrino emission. Although this method does not place rigorous limits, it does find that the neutrino flux from thermal-relic DM annihilation can rival that inferred for SM emission templates.  
\end{itemize} 
The former uses the spatial distribution of events to place a robust and conservative constraint that does not rely on any specific Galactic CR emission model; the latter uses the template-dependent Inner Galaxy neutrino intensities inferred by IceCube to assess the DM contribution under different assumptions about Galactic neutrino emission. 

The conservative spatial analysis alone improves existing limits from IceCube and ANTARES by factors of a few over the $\sim$ a few TeV – a few $100\ {\rm TeV}$ mass range, with still stronger constraints obtained when the Galactic CR contribution is subtracted. These results demonstrate that measurements of the Galactic neutrino sky have now entered the regime in which they can directly test the thermal relic hypothesis for DM annihilating into neutrinos.

\begin{figure*}
\centering
\hspace{-0.2cm}
\includegraphics[width=0.49\textwidth]{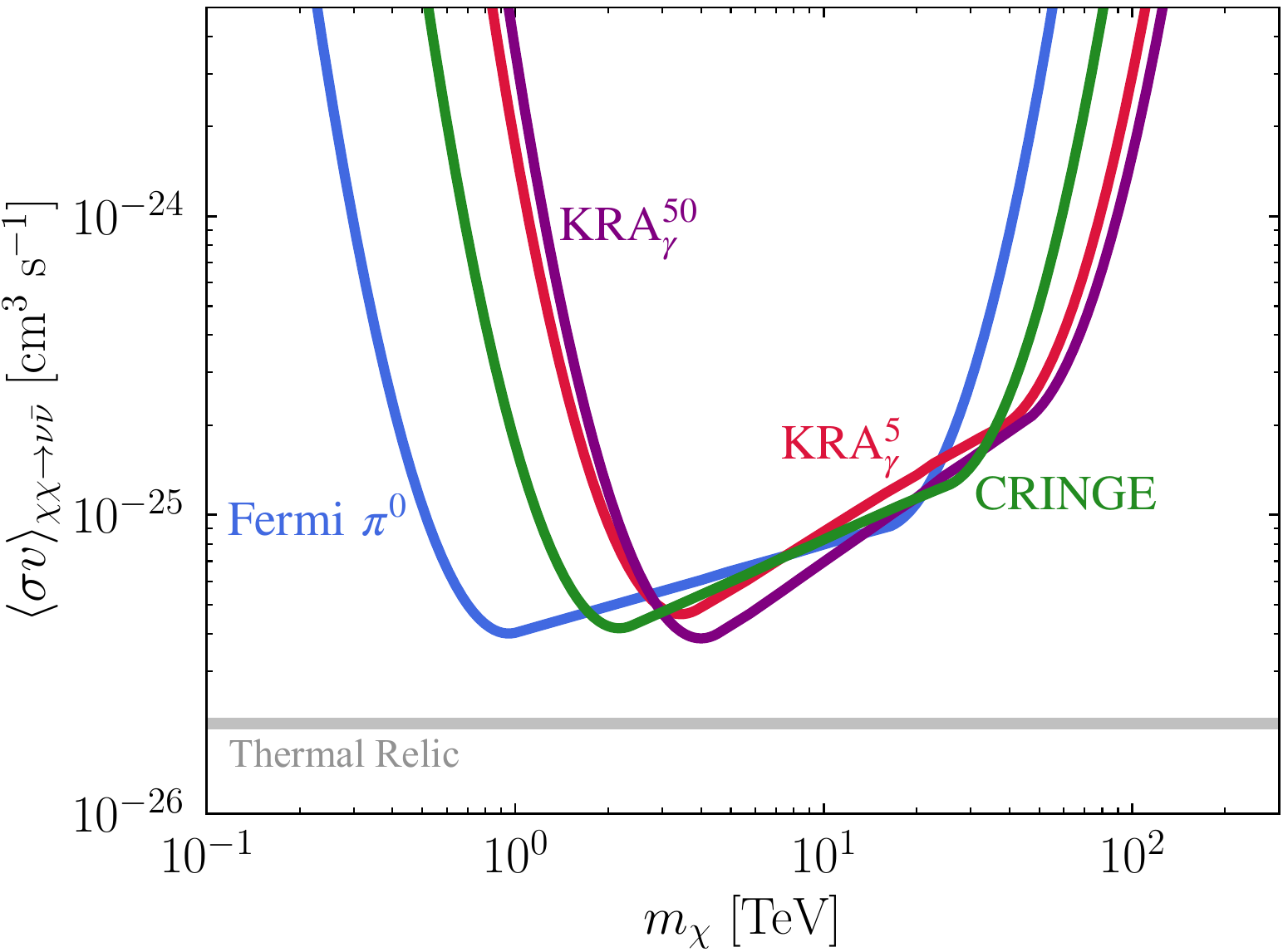} \ \ 
\includegraphics[width=0.49\textwidth]{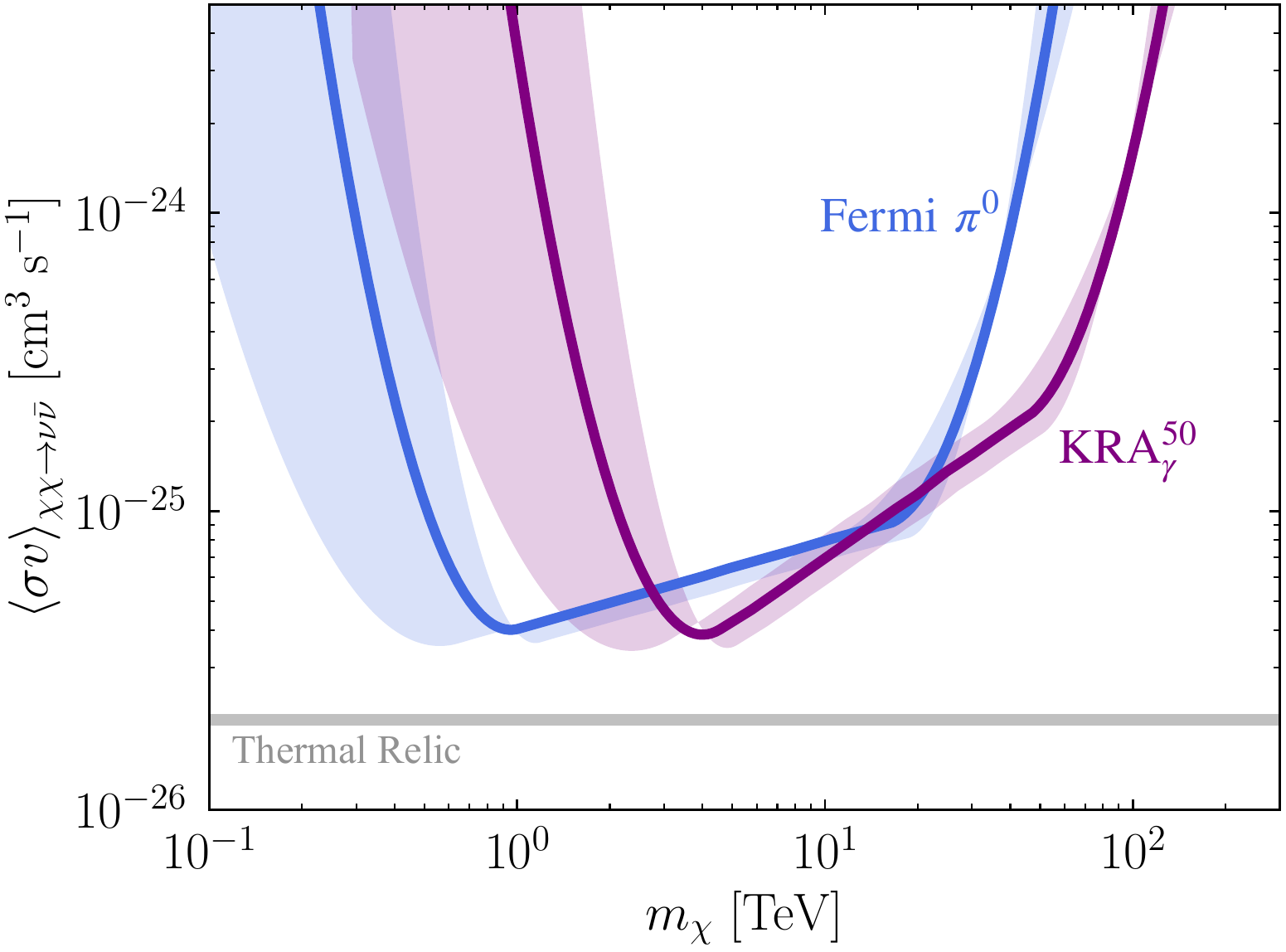}
\caption{\label{fig:ann_lim_envelope} Spectral-envelope comparison of the neutrino intensity from $\chi\chi\to\bar\nu\nu$ annihilation with the IceCube best-fit Inner Galaxy (IG) neutrino intensity using Eq.~\eqref{eq:flux-comparison} with the neutrino line-smearing prescription from Eq.~\eqref{eq:dNdE_log}. \emph{Left:} Curves for the four Galactic CR templates considered by IceCube. For each DM mass, the cross section is chosen such that the DM-induced intensity reaches the corresponding best-fit template intensity, $\Phi_\nu^\chi=\Phi_{\nu}^{T}$, at one energy within the support of the template. All curves shown here adopt the fiducial value $\sigma_{\log_{10} E} = 0.20$ for the energy spectrum in Eq.~\eqref{eq:dNdE_log}. \emph{Right:} Dependence of the comparison on the width
$\sigma_{\log_{10}\!E}$ of the log-normal DM neutrino spectrum, shown for the Fermi $\pi^0$ and KRA$_\gamma^{50}$ templates. Solid curves denote the fiducial choice $\sigma_{\log_{10}\! E}=0.20$, while the shaded bands span $\sigma_{\log_{10}\!E}=0.15$ - $0.30$. The modest variation illustrates that the spectral-envelope comparison is only weakly dependent on the assumed line width. Note that these curves are not statistical limits, but illustrate the level of DM annihilation intensity that is comparable to the Galactic neutrino emission measured by IceCube.
}
\end{figure*}

\section{Formalism}
Dark matter annihilation via $\chi\chi \to \nu \bar \nu$ in the Galactic halo can produce an observable neutrino flux. In this work, we consider self-conjugate DM undergoing velocity-independent ($s$-wave) annihilation into neutrino-antineutrino pairs characterized by the velocity-averaged annihilation cross section $\langle \sigma v \rangle$. The differential neutrino flux can be written~\cite{Arguelles:2019ouk}
\be
\label{eq:dmnu_flux}
\frac{d\phi_\nu}{dE_\nu d\Omega}  = \frac{\langle\sigma v\rangle} {8\pi m_\chi^2} \frac{dN_\nu}{dE_\nu} \bar J\,,
\ee
where $m_\chi$ is the mass of the DM particle, $dN_\nu/dE_\nu$ is the neutrino spectrum produced per annihilation, and $\bar J$ is the directional $J$-factor 
\begin{equation}
\bar J(\Omega) \equiv \int_{\rm LOS} ds\, \rho_\chi^2 [r(s,\ell,b)]\,,   
\end{equation}
where $s$ is a line-of-sight (LOS) coordinate,   
$\ell$ and $b$ are the Galactic longitude and latitude respectively, 
and we have defined 
\begin{equation}
r \equiv   \sqrt{R_\odot^2 + s^2 - 2R_\odot s \cos \ell \cos b}  ~,  
\end{equation}
as the Galactocentric distance of a point located at a distance $s$ along the line of sight, where $R_\odot = 8.2$ kpc 
is the solar distance from the Galactic Center \cite{McMillan_2016}. 
We model the halo as a NFW profile~\cite{Navarro:1995iw}, 
\be
\rho_{\chi} (r) = \frac{ \rho_s}{  (r/r_s) \left(1+ r/r_s \right)^2},
\ee
where  $r_s = 20$ kpc is the scale radius and $\rho_s$ is the scale density normalized to match the local DM density $\rho_\chi(R_\odot) = 0.4\ {\rm GeV/cm^{3}}$ \cite{S_ding_2025}.

Since our spectral analysis considers an extended region of the Inner Galaxy, the differential neutrino flux averaged over a region of interest (ROI) is 
\be
\label{eq:Phi_chi}
\Phi_\nu^{\chi} (E_\nu)  
\equiv 
\frac{1}{\Delta \Omega} \int_{\Delta \Omega} d\Omega \left( \frac{d\phi_\nu}{ dE_\nu d\Omega  } 
\right),
\ee
where
$d\Omega = \cos b\, db\, d\ell$, $\Delta\Omega$ is the solid angle over the ROI, and the integrand is given in Eq.~\eqref{eq:dmnu_flux}.  After evaluating the angular integral, the resulting flux is proportional to the $J$-factor 
\be
\label{eq:j_int_roi}
J = \int_{\Delta \Omega} d\Omega\, \bar J =
\int_{\Delta \Omega} d\Omega \int_{\rm LOS}  ds\, \rho_\chi^2[r(s,\ell,b)],
\ee
where, for an ROI with $|\ell| \le 20^\circ$ and $|b| \le 15^\circ$, we obtain $J = 6.6 \times 10^{22}$ GeV$^2$cm$^{-5}$.

For non-relativistic DM, $\chi\chi \to \bar \nu \nu$ annihilation yields a monochromatic spectrum with $E_\nu = m_\chi$. Assuming standard three-flavor neutrino oscillations during propagation through the Galactic halo, the neutrino energy spectrum per DM annihilation is
\be
\label{eq:nu_spec}
\frac{dN_\nu}{dE_\nu} = \frac{2}{3} \delta (E_\nu - m_\chi)\,,
\ee
where the factor of $2$
accounts for the neutrino-antineutrino pair produced in each annihilation, while the factor $1/3$ reflects the approximately $1:1:1$ flavor composition at Earth after oscillations.

\section{Neutrinos from the Galactic Plane}
The Galactic Plane is a guaranteed source of high-energy neutrinos produced primarily through hadronic interactions of Galactic CRs with the interstellar medium (ISM). These interactions produce secondary mesons, in particular $\pi^0$ and $\pi^\pm$, where $\pi^0$ decays to produce high-energy gamma rays, while the decay chain of $\pi^\pm$ produces neutrinos and antineutrinos. The resulting diffuse gamma ray and neutrino emission is therefore determined by the underlying Galactic CR population and its transport through the interstellar medium, leading to different predictions for the spatial and spectral distribution of the diffuse Galactic neutrino emission.
Searches for this diffuse high-energy neutrino emission have been carried out by various high-energy neutrino telescopes like IceCube~\cite{IceCube:2023ame}, ANTARES~\cite{ANTARES:2022izu}, and KM3NeT~\cite{KM3NeT:2026omo}.
More recently, using an enlarged dataset together with improved detector calibration, ice modeling, and event reconstruction, IceCube established high-energy neutrino emission from the Milky Way at a post-trial significance of $5.7\sigma$~\cite{IceCube:2026plr}. The observed diffuse emission is spatially extended along the Galactic Plane and is strongly concentrated toward the Inner Galaxy, making it a particularly sensitive probe of any additional contribution from DM annihilation.

The latest IceCube analysis combines events from three different topologies: showers, starting tracks, and through-going tracks, and performs a combined likelihood analysis, with the sensitivity to the Inner Galaxy dominated by the shower sample. The diffuse Galactic emission is modeled using four representative CR templates:
\begin{itemize}
\item \textbf{Fermi-LAT $\boldsymbol{\pi^0}$:}
This template is based on the morphology of diffuse $\gamma$-ray emission and assumes a spatially uniform CR diffusion scenario resulting in a neutrino spectrum $\propto E_\nu^{-2.7}$
\cite{Fermi-LAT:2012edv}.
\item \textbf{\bf KRA\boldsymbol{$_\gamma^5$}:}
This  model incorporates spatially dependent CR transport and predicts a harder spectrum with a more centrally concentrated morphology ~\cite{Gaggero:2015xza}. The KRA$_\gamma^5$ variant corresponds to the maximum rigidity of the CRs set to $5$ peta-Volts (PV).
\item \textbf{\bf KRA\boldsymbol{$_\gamma^{50}$}:}
Similar to the KRA$_\gamma^5$ model, the KRA$_\gamma^{50}$ variant has the maximum rigidity of the CRs set to $50$ PV~\cite{Gaggero:2015xza}.

\item {\bf CRINGE:}
This model is obtained from a global fit to the CR measurements and includes an unresolved Galactic-source contribution, yielding a spatial and spectral distribution intermediate between the Fermi $\pi^0$ and KRA$_\gamma$ models \cite{Schwefer:2022zly}.
\end{itemize}
Throughout this work, we use these templates and perform various analyses to constrain the $\chi\chi \to \bar \nu \nu$ annihilation cross section. 

\section{Longitude Distribution Analysis}
In Ref.~\cite{IceCube:2026plr} IceCube presents the background-subtracted distribution of shower events as a function of reconstructed Galactic longitude $\widehat{\ell}$. Shower events with reconstructed energy $\widehat{E}_\nu > 5\ {\rm TeV}$ and reconstructed Galactic latitude $|\widehat{b}|<15^\circ$ are selected, while the background is estimated from the off-plane data. The resulting background-subtracted event distribution is reported in $40^\circ$ bins of $\widehat{\ell}$ and is therefore independent of the assumed Galactic emission template. In the central bin, $|\widehat{\ell}| < 20^\circ$, IceCube observes $217$ events over an estimated background of $154.4 \pm 4.1$, corresponding to an excess of $62.6 \pm 15.3$ events. This data-driven event excess provides an independent observable with which we compare the DM signal prediction. 

\medskip

{\bf Model Independent Spatial Analysis.}
The most conservative analysis criterion for constraining DM annihilation compares the DM contribution directly with the IceCube background-subtracted longitude distribution. Importantly, this analysis does not subtract any contribution from Galactic neutrino emission and therefore does not assume any additional contribution from SM sources presumed to be active in the Galactic Plane. Thus, this approach is model independent and does not rely on any templates mentioned in the previous section.

For a given reconstructed longitude bin $i$ and fixed $m_\chi$, we account for the reconstructed energy selection and forward-fold the DM signal through the shower angular response, applying the same reconstructed spatial and energy selections used by IceCube for the residuals. The expected number of selected DM-induced shower events reconstructed in the on-plane region is then given by
\be
\label{eq:N_i}
\! N^{\chi, \rm on}_i = \tau  \! \int d\Omega \! \int_0^\infty \! \! dE_\nu 
\frac{d\phi_\nu}{dE_\nu d\Omega} A_{\rm eff}(E_\nu,\delta) \varepsilon_i (E_\nu, \Omega), 
 \! \!
\ee
where the differential flux is from Eq.~\eqref{eq:dmnu_flux}, $\tau = 12$ years is the IceCube livetime \cite{IceCube:2026plr}, $A_{\rm eff}$ is the flavor-summed shower effective area and is given by
\be
\!  \! A_{\rm eff}(E_\nu,\delta) = A^N_{\rm eff}(E_\nu) \Theta(\delta +  5^\circ)
+ 
A^S_{\rm eff}(E_\nu) \Theta(-5^\circ - \delta) ,\! \! 
\ee
where $\Theta$ is the Heaviside theta function, $\delta(\ell, b)$ is the declination and the $N,S$ terms correspond to northern and southern declination regions, respectively \cite{IceCube:2026plr}. Here, the angular integration in Eq.~\eqref{eq:N_i} is over the full true sky ($|\ell| < 180^\circ$ and $|b|< 90^\circ$) and we have defined 
\be
\label{eq:epsi}
\varepsilon_i(E_\nu,\Omega) \equiv \varepsilon_5(E_\nu) \varepsilon_i^{\rm \rm ang} (E_\nu, \Omega)~,
\ee
where $\varepsilon_5$ accounts for the probability that a neutrino with true energy $E_\nu$ gets reconstructed with energy proxy $\widehat{E} > 5$ TeV, 
\be
\label{eq:eps_reco}
\varepsilon_5 (E_\nu) \equiv {\cal P}_E(\widehat{E} > 5\ {\rm TeV}| E_\nu)\,,
\ee
where ${\cal P}_E$ is the IceCube energy-reconstruction probability distribution taken from  Ref.~\cite{IceCube:2023ame} (see End Matter).
Similarly,  $\epsilon^{\rm ang}_i$
which accounts for the probability that an event arriving from the true direction $(\ell,b)$ will be reconstructed within the $i^\text{th}$ longitude bin and Galactic Plane selection $|\widehat{b}| < 15^\circ$,
\be
\label{eq:pang}
\varepsilon^{\rm ang}_i(E_\nu, \Omega) \equiv {\cal P}_\Omega \left(\widehat{\ell} \in \Delta{\widehat {\ell}_i} \, , \, |\widehat{b}|<15^\circ  \middle| \ell, b, E_\nu \right)\,,
\ee
where ${\cal P}_\Omega$ is the IceCube angular-reconstruction probability distribution taken from Ref.~\cite{IceCube:2023ame} (see End Matter).
Thus, the full true-sky information is included for the event prediction in each bin, while the spatial selection is imposed only on reconstructed angular coordinates. 

Since the IceCube residuals are obtained using a data-driven off-plane background estimate, DM events reconstructed in the off-plane control region also contribute to the subtracted background. We therefore apply the same on-off subtraction to the predicted DM distribution and denote the resulting DM residual in the reconstructed $i^{\rm th}$ longitude bin as
\be
N^{\chi,\rm res}_i = N_i^{\chi,\rm on} - N_i^{\chi,\rm off \rightarrow \rm bkg}\,,
\ee
where the second term includes the same declination-dependent solid angle rescaling used in the IceCube background estimate. Details of the reconstructed energy efficiency, angular reconstruction, and data-driven background subtraction are provided in the End Matter.

To constrain $\langle\sigma v\rangle$ for each value of $m_\chi$, we perform a $\chi^2$ test. 
The number of observed background-subtracted shower events in each bin is $N_{i}^{\rm obs}$, so 
\be
\label{eq:chi2}
\chi^2(\langle \sigma v\rangle) = \sum_i \frac{(N_i^{\rm obs} - N_i^{\chi,\rm res})^2}{\sigma_i^2},
\ee
where the sum is over all reconstructed longitude bins, $\sigma_i$ is the reported uncertainty on the measured residual of the $i^{\rm th}$ bin, and $N_i^{\chi,\rm res}$ implicitly depends on $m_\chi$ and $\langle \sigma v\rangle$. 
For each $m_\chi$ we define  $\chi^2_{\min} \equiv \min[ \chi^2 \big( \langle\sigma v\rangle \big)]$ for ${\langle\sigma v\rangle\geq 0}$, and the upper-limit on $\langle \sigma v \rangle$ satisfies
\be 
\label{eq:chi2-criterion}
\chi^2(\langle\sigma v\rangle_{\rm lim}) - \chi^2_{\rm min} = \Delta \chi^2_{\rm CL},
\ee
where $\Delta \chi^2_{\rm CL}$ is the threshold corresponding to the desired confidence level (CL), and we  choose $\Delta \chi^2_{\rm CL} = 2.71$, corresponding to the $90\%$ CL upper-limit on $\langle \sigma v \rangle$. In Fig. \ref{fig:ann_lim_chisq}, the thick orange curve labeled ``IceCube Galactic Plane" shows the  constraint we derive using this conservative method, which does not rely on any CR model of Galactic neutrinos to set these limits.

\medskip

{\bf Template-Subtracted Spatial Analysis.}
Although the previous model-independent analysis is highly robust, it ignores any potential contributions from Galactic neutrino emission. 
To assess the impact of specific models, we repeat our spatial analysis after subtracting the corresponding Galactic CR contribution scaled to its IceCube best-fit normalization from the background-subtracted shower event residuals. In this analysis, we follow the same procedure outlined in Eqs.~\eqref{eq:N_i} - \eqref{eq:chi2-criterion},
however the residuals in Eq.~\eqref{eq:chi2} are replaced by the template-subtracted residuals for each CR model,
\be
N_i^{\rm obs} \to N_i^{\rm obs} - N_i^{T},
\ee
where $N_i^T$ is the expected number of events in the $i^{\rm th}$ reconstructed longitude bin for template $T$, scaled to the best-fit normalization reported by IceCube, where $T = \{$Fermi $\pi^0$, KRA$_\gamma^5$, CRINGE$\}$.

In this analysis, part of the observed Galactic emission is attributed to astrophysical neutrinos, reducing the residual emission against which the DM contribution is constrained and consequently strengthening the limits on $\langle \sigma v \rangle$.
In Fig.~\ref{fig:ann_lim_chisq}, we show our 90\% CL limits derived using the spatial residuals binned in longitude. We find that subtracting the Galactic CR templates further strengthens the limits by a factor of a few (colored curves) compared to the model-independent analysis (thick orange curve).
    
\section{Template Energy Spectrum Analysis}
Alternatively, we use the diffuse per-flavor ($\nu+\bar \nu$) intensity within the region $|\ell|<20^\circ$ and $|b|<15^\circ$, inferred by IceCube for each of the four Galactic CR emission templates. For each template, the predicted Inner Galaxy (IG) spectrum is scaled by the best-fit normalization obtained from the global IceCube likelihood analysis.
Although the underlying CR models predict substantially different morphologies and spectra, the corresponding best-fit neutrino intensities inferred by IceCube converge to similar values over the energy range where the analysis is most sensitive.

We compare the predicted neutrino intensity from DM annihilation with the intensity inferred by IceCube for each Galactic CR template.
Specifically, for each $m_\chi$ we determine the largest $\langle\sigma v \rangle$ satisfying
\be
\label{eq:flux-comparison}
\Phi_\nu^{\chi}(E_\nu) \le 
\Phi^T_{\nu}(E_\nu) ~,~ \forall \, E_\nu \in [E_{\nu,T}^{\rm min}, E_{\nu,T}^{\rm max}],
\ee
where $\Phi_\nu^\chi$ is given in Eq.~\eqref{eq:Phi_chi} and $\Phi^{T }_{\nu}$ denotes the best-fit intensity for $T \in \{{\rm Fermi\ \pi^0, KRA_\gamma^5, KRA_\gamma^{50}, CRINGE}\}$, and the energy range  $[E_{\nu,T}^{\rm min}, E_{\nu,T}^{\rm max}]$ that IceCube is sensitive to differs slightly for each  $T$ \cite{IceCube:2026plr}. Here both $\Phi_\nu^\chi$ and $\Phi_{\nu}^T$ are averaged over the Inner Galaxy ROI defined by $|\ell|<20^\circ$ and $|b|<15^\circ$. This comparison conservatively allows the entire best-fit IG flux to originate from DM annihilation.

While the physical neutrino spectrum is monochromatic (see Eq.~\ref{eq:nu_spec}), for the spectral comparison we represent the $\delta$-function numerically by a normalized log-normal distribution such that
\be
\label{eq:dNdE_log}
\frac{dN_\nu}{dE_\nu} =  \frac{2/3}{\sqrt{2\pi} \sigma_{\ln E} E_\nu} \exp \left[ -\frac{1}{2} \left( \frac{\ln \big( E_\nu/m_\chi \big)}{\sigma_{\ln E}} \right)^2 \right],
\ee
with $\sigma_{\log_{10}\!E}=0.2$, that is, $\sigma_{\ln E} = (\ln 10) \sigma_{\log_{10} E} \simeq 0.46$. This finite width provides a smooth numerical representation of the monochromatic injection spectrum and is not intended to model detector-level energy resolution. 

In the \emph{left} panel of Fig.~\ref{fig:ann_lim_envelope}, we show the spectral-envelope comparison defined by Eq.~\eqref{eq:flux-comparison} for the different Galactic CR emission models. Remarkably, the corresponding cross sections approach the benchmark thermal relic value. Since the best-fit spectra for different Galactic CR models span slightly different energy ranges, the smallest $\langle \sigma v \rangle$ occur at somewhat different DM masses. At DM masses outside the energy range covered by a given best-fit spectrum, a larger annihilation cross section is required for the DM intensity to reach the corresponding spectral envelope, which causes the curves in this figure to rise for both low and high masses. Consequently, the Fermi $\pi^0$ template, which extends down to $\sim 1$ TeV energies, provides the greatest sensitivity around $1 - 2$ TeV, whereas the KRA$_\gamma$ models achieve their greatest sensitivity at masses of $\mathcal{O}(10)$ TeV, assuming the fiducial value of $\sigma_{\log_{10}\! E} = 0.20$.

The \emph{right} panel of Fig.~\ref{fig:ann_lim_envelope} illustrates the dependence of this comparison on the numerical width of the monochromatic line. For the representative Fermi $\pi^0$ and KRA$_\gamma^{50}$ templates, we vary $\sigma_{\log_{10} E}$ from $0.15 - 0.30$ around the fiducial value of $0.2$. The resulting variation in the spectral envelope curves is modest, showing that the comparison is not strongly dependent on the precise width adopted to represent the monochromatic spectrum. We emphasize that the curves in Fig.~\ref{fig:ann_lim_envelope} are not statistical limits. Rather, they identify the annihilation cross section for which the predicted DM neutrino intensity becomes comparable to the best-fit Galactic neutrino intensity inferred by IceCube. The proximity of these curves to $\langle \sigma v \rangle_{\rm th}$ demonstrates that the measured IG neutrino intensity is already comparable to that expected from the thermal-relic DM annihilation over part of the TeV mass range.

\section{Comparison with existing limits}
The existing limits from IceCube and ANTARES are shown in Fig.~\ref{fig:ann_lim_chisq}. For IceCube, we show the $90\%$ CL limits obtained using 5 years of DeepCore data, which constrains neutrino signals from DM annihilation in the Galactic Center~\cite{IceCube:2023ies}. For ANTARES,\footnote{More recent preliminary constraints using the full ANTARES data set have also been presented in Ref.~\cite{Gozzini:2023nka}, which we do not display here since these only appear in an unpublished conference note.} we show the published $90\%$ CL constraints from the dedicated Galactic Center searches of Refs.~\cite{ANTARES:2015vis,Albert:2016emp}, following the combined presentation of Ref.~\cite{Arguelles:2019ouk}. These searches assume a NFW DM halo profile and search for direct annihilation into neutrino-antineutrino pairs. Constraints based on recent KM3NeT/ARCA observations have also been independently derived~\cite{KM3NeT:2024xca}.

Even the robust, model-independent residual analysis presented here improves upon the existing constraints over much of the mass range from a few TeV - a few $100$ TeV. The improved sensitivity relative to the earlier IceCube DeepCore (DC) search can be understood primarily from the substantially larger multi-TeV shower acceptance of the present IceCube event sample, together with the longer exposure. From the published effective areas, the present shower selection provides roughly an order-of-magnitude larger effective acceptance in the several-TeV range after accounting for the reconstructed-energy threshold, while the 12-year data set provides an additional factor of $12/5 \simeq 2.4$ in livetime. The larger effective area reflects the use of the full IceCube detector for the high-energy shower sample, whereas the earlier search employed a contained-event selection with DeepCore as the fiducial volume~\cite{IceCube:2016oqp}. These gains are partially offset in our analysis by the reconstructed-energy selection and by the data-driven background subtraction, which removes part of the spatially extended DM signal.

The model-dependent residual analysis strengthens these limits further, reaching improvements of roughly an order of magnitude over the existing constraints in part of the relevant mass range explored in this work. This improvement is driven by the recent detection of diffuse Galactic neutrino emission by IceCube, which provides a powerful new constraint on any additional contribution from DM annihilation to the observed Galactic signal.

\section{Conclusions \& Outlook}
In this \emph{Letter}, we have shown that the recent high-significance detection of diffuse high-energy neutrino emission from the Galactic Plane by IceCube provides a powerful new probe of DM annihilation to neutrinos.
Using the background-subtracted reconstructed Galactic longitude distribution of shower events, we derived new constraints on $\chi \chi \rightarrow \nu \bar{\nu}$, while the template-dependent Inner Galaxy neutrino spectra provide a complementary assessment of the sensitivity to a DM contribution. The limits obtained from the spatial distribution alone are already more stringent than previous constraints from IceCube and ANTARES over much of the $ 1- 100$ TeV mass range, while the spatial template-subtracted analyses probe even smaller cross sections, nearly approaching thermal-relic parameter values.

Our current work is conservative in several respects and can be extended in a number of directions. Our spatial constraints are derived using only the shower event sample, but incorporating the track events in a combined morphological analysis should further improve the sensitivity. Likewise, while we have focused on the monochromatic neutrino signal from direct annihilation into $\nu \bar{\nu}$, realistic DM models generally also produce secondary neutrinos from the decays of $W$ and $Z$ bosons, charged leptons, and hadrons, providing additional channels that can be constrained within the same framework. Furthermore, the substantially improved statistics and angular resolution expected from IceCube-Gen2~\cite{IceCube-Gen2:2020qha,Gen2_TDR} and KM3NeT~\cite{KM3Net:2016zxf} will enable increasingly precise measurements of diffuse Galactic neutrino emission, significantly extending the sensitivity to DM annihilation.

\acknowledgments
\textit{\textbf{Acknowledgments.}}
We are particularly grateful to Dan Hooper for clarifying important subtleties about this analysis. We also thank Joshua Foster, Matheus Hostert,  Shirley Weishi Li, Pedro Machado, Alex Drlica-Wagner, and Bei Zhou for feedback on the manuscript.
We are thankful for the Cosmic Physics Center MUNCH Journal Club at Fermilab where this idea originated.
We acknowledge support from the FermiForward Discovery Group, LLC under Contract No. 89243024CSC000002 with the U.S. Department of Energy, Office of Science, Office of High Energy Physics. M.\,M. also acknowledges the support of the NSF-Simons AI-Institute for the Sky (SkAI) via grants NSF AST-2421845 and Simons Foundation MPS-AI-00010513.
\bibstyle{apsrev4-2}
\bibliography{refs}
\newpage
\onecolumngrid
\bigskip
\bigskip
\begin{center}
\textbf{\large End Matter}
\end{center}
\bigskip
\twocolumngrid
\section{ $J$-factor longitude dependence}
\begin{figure}
\centering
\includegraphics[width=0.45\textwidth]{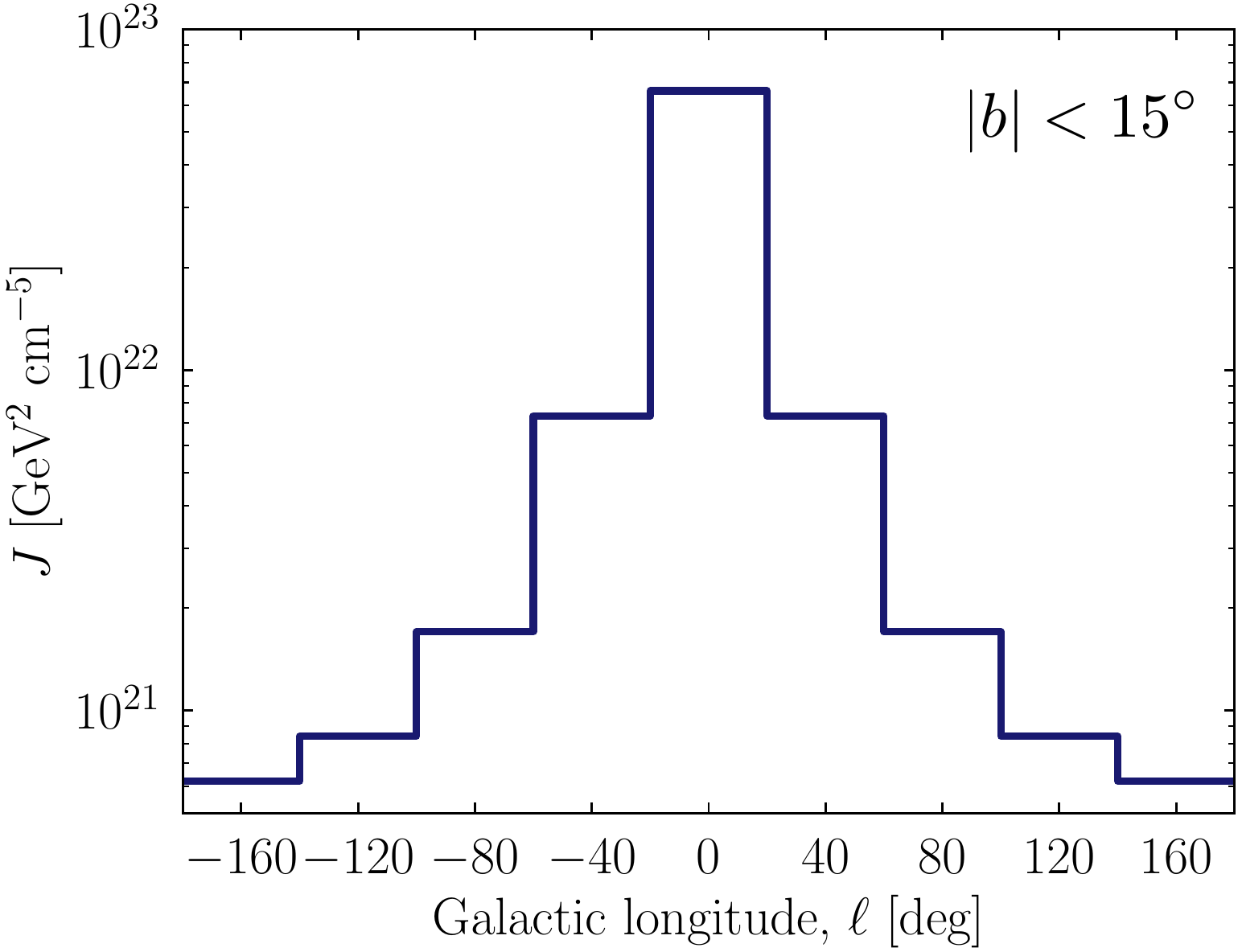}
\caption{\label{fig:jfact_longi}Galactic longitude dependence of the annihilation $J$-factor defined in Eq.~\eqref{eq:j_int_roi} for the NFW halo profile. The integration is performed over $40^\circ$ longitude bins with $|b| < 15^\circ$, illustrating the intrinsic Galactic-longitude dependence of the DM annihilation signal.
}
\end{figure}

In Fig.~\ref{fig:jfact_longi}, we show the annihilation $J$-factor defined in Eq.~\eqref{eq:j_int_roi} as a function of true Galactic longitude $\ell$. The $J$-factor is computed in $40^\circ$ bins of $\ell$, with the integration in each bin performed over true Galactic latitude $|b| < 15^\circ$. The corresponding solid angle is
\be
\Delta \Omega_i= \Delta \ell_i \big[ \sin 15^\circ - \sin (-15^\circ) \big] = 0.361\ {\rm sr}\,,
\ee
where $\Delta \ell_i = 40^\circ$. The resulting $J$-factor is strongly peaked toward the Galactic Center owing to the steep rise of the NFW DM density. This central concentration sets the underlying true-sky morphology of the DM signal. The DM event distribution used in the spatial analysis additionally accounts for shower angular reconstruction, the reconstructed spatial selection, and the data-driven background subtraction, as shown in Fig.~\ref{fig:residual_vis}.

\section{Additional details on the longitude-based spatial constraints}

\begin{figure}
\centering
\includegraphics[width=0.49\textwidth]{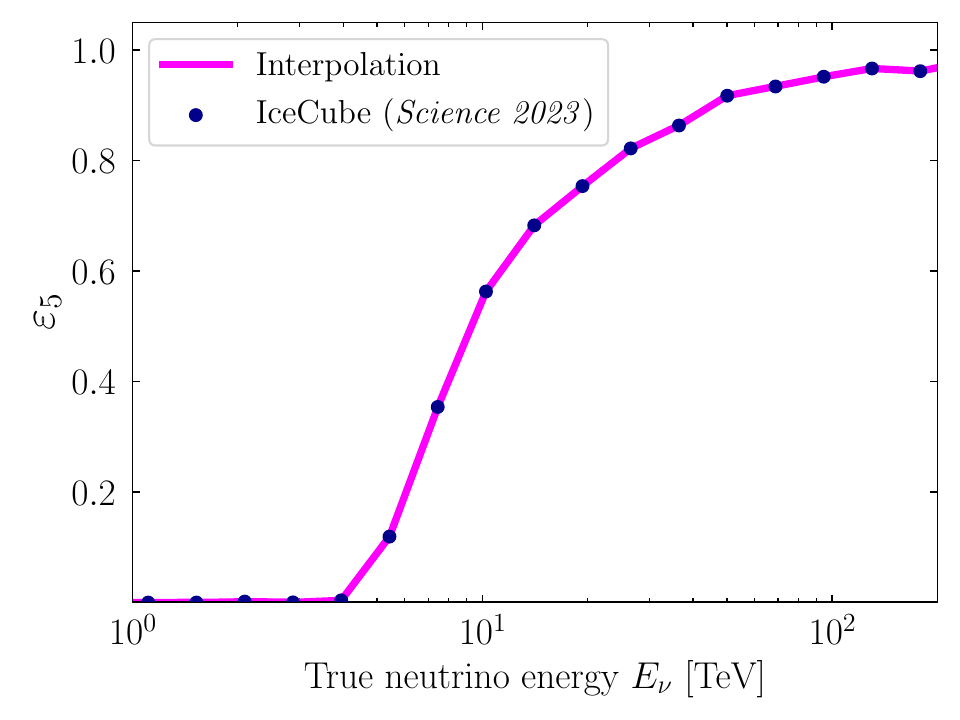}
\caption{\label{fig:epsilon5tev} Reconstructed-energy threshold efficiency $\epsilon_5(E_\nu)\equiv P(\widehat E > 5\ {\rm TeV}\mid E_\nu)$ as a function of true neutrino energy $E_\nu$. The points are obtained by integrating the publicly released IceCube Monte Carlo energy-response distribution underlying Fig. S6A of Ref.~\cite{IceCube:2023ame} above $\widehat E_\nu=5\ {\rm TeV}$.
}
\end{figure}
\textit{\textbf{Energy reconstruction of shower events.}} For an appropriate comparison with the residuals from Ref.~\cite{IceCube:2026plr}, the shower-event longitude distribution used in our spatial analysis is subject to the same reconstructed energy requirement, $\widehat{E} > 5\ {\rm TeV}$. The effective areas, on the other hand, are provided in terms of true neutrino energy $E_\nu$ in Eq.~\eqref{eq:N_i}. However, the corresponding reconstructed shower energy response for the shower sample using the DNNC selection criteria is not publicly available. Therefore, we approximate this response using the publicly released Monte Carlo energy-response distribution from the previous IceCube Galactic Plane neutrino analysis from Ref.~\cite{IceCube:2023ame} (see Fig. S6A).  This distribution provides the reconstructed shower energy proxy $\widehat{E}$ as a function of true neutrino energy $E_\nu$ for the simulated shower sample.

Given this distribution, we construct the threshold efficiency from Eq.~\eqref{eq:eps_reco},
\be
\!\!
\varepsilon_5 (E_\nu) \equiv {\cal P}_E(\widehat{E} > 5\ {\rm TeV}| E_\nu) = \int_{5\ {\rm TeV}}^\infty \! d \widehat{E} \ p(\widehat{E} | E_\nu)\,, 
\! \!
\ee
where $p$ is the conditional probability density of the reconstructed shower energy $\widehat{E}$ given a true neutrino energy $E_\nu$. We show the variation of $\varepsilon_5$ with respect to $E_\nu$ in Fig.~\ref{fig:epsilon5tev}. Therefore, we do not impose a cut on $E_\nu$ but the migration across the $\widehat{E} =5\ {\rm TeV}$ threshold is instead incorporated through $\varepsilon_5(E_\nu)$. The updated IceCube analysis reports substantial improvements in shower directional reconstruction relative to the preceding analysis, but does not provide an updated reconstructed-energy response, so we use the published response of Ref.~\cite{IceCube:2023ame} as a conservative approximation.

\begin{figure*}
\centering
\includegraphics[width=0.49\textwidth]{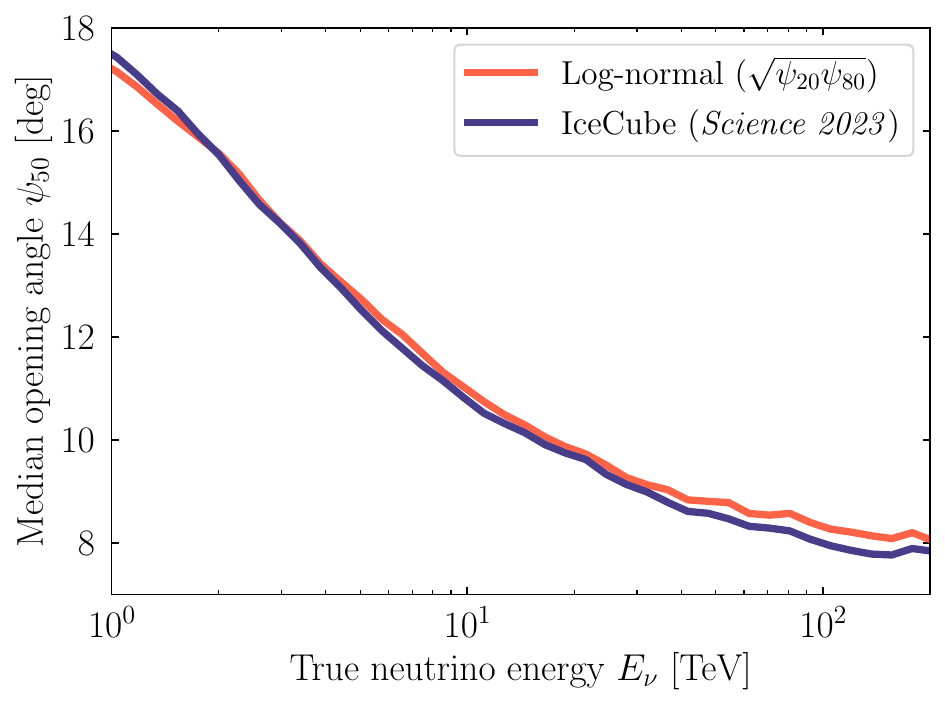}
\includegraphics[width=0.49\textwidth]{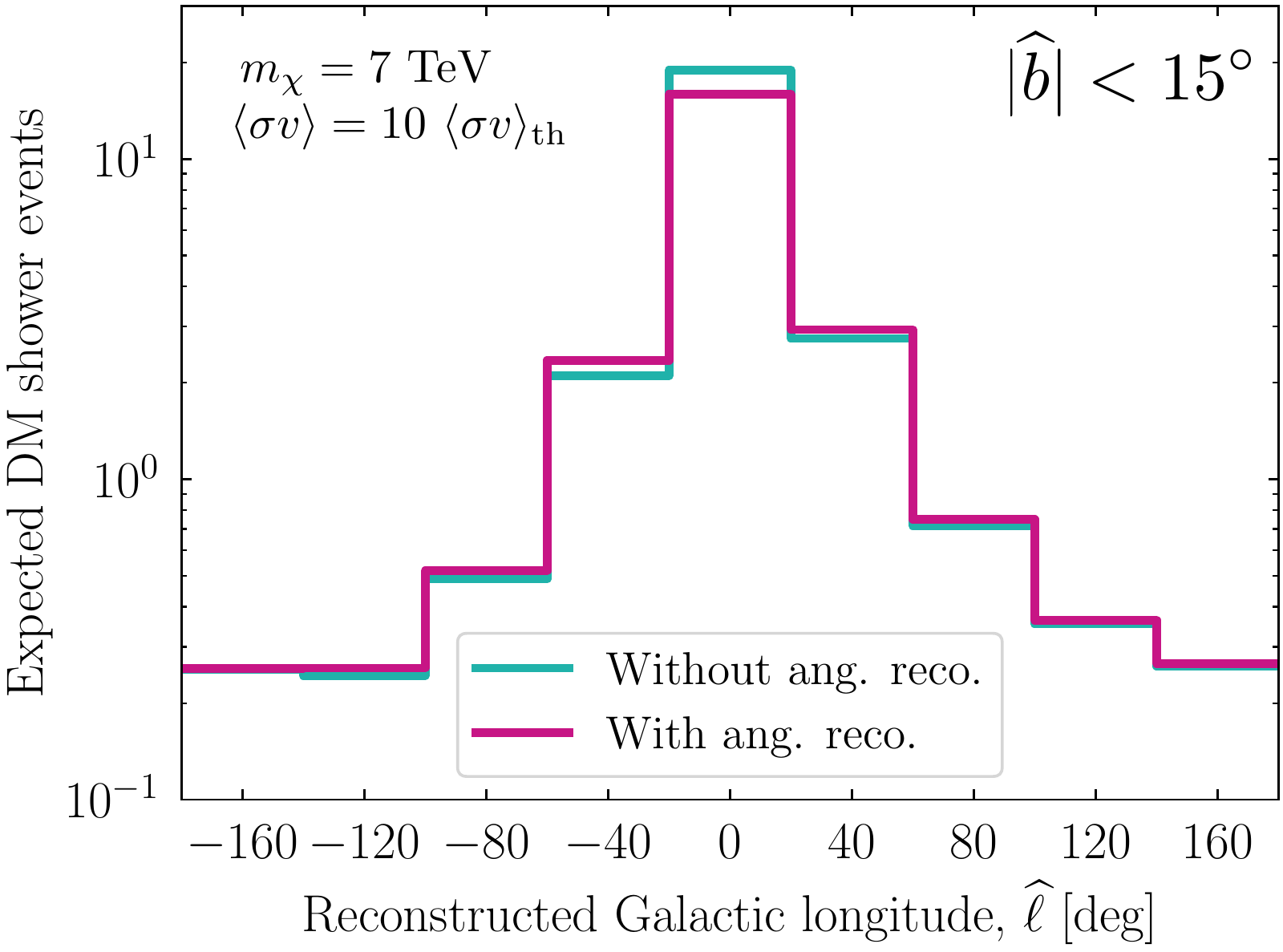}
\caption{\label{fig:ang_recon}Angular reconstruction of shower events. \emph{Left: }Median opening angle $\psi_{50}$ from the publicly released IceCube angular-response information of Ref.~\cite{IceCube:2023ame}, compared with the log-normal prediction $\psi_{50}=\sqrt{\psi_{20}\psi_{80}}$. \emph{Right: }Expected DM shower-event distribution in reconstructed Galactic longitude with (pink) and without (sea-green) forward folding through the shower angular response, illustrating the resulting modification of the spatial morphology. The forward-folded distribution is selected using $|\widehat{b}| < 15^\circ$. Note that the curve corresponding to without angular reconstruction is the perfect reconstruction limit where $(\widehat{\ell},\widehat{b}) = (\ell,b)$.
}
\end{figure*}
\textit{\textbf{Angular reconstruction of shower events.}} The shower event residuals used in our spatial analysis are reported by IceCube as a function of reconstructed Galactic coordinates $(\widehat{\ell},\widehat b)$, whereas the DM intensity in Eq.~\eqref{eq:dmnu_flux} is specified in terms of the true direction $(\ell,b)$. We therefore forward-fold the predicted DM signal through the shower angular response before applying the spatial selection. The event-level angular response information required to reproduce the spatial reconstruction in Ref.~\cite{IceCube:2026plr} is obtained from IceCube detector simulations, which are not publicly available. We instead use the publicly released angular-resolution information from the preceding IceCube Galactic Plane 
analysis in Ref.~\cite{IceCube:2023ame}.

For an individual event, we define the angular reconstruction error by the opening angle $\Psi$
\be
\Psi = \cos^{-1} \big( \widehat{\mathbf n}\cdot{\mathbf n} \big)\,,
\ee
where $\mathbf{n}$ and $\widehat{\mathbf{n}}$ are unit vectors along the true and reconstructed neutrino directions, respectively and $0 \leq \Psi \leq \pi$. At a fixed true neutrino energy $E_\nu$, the reconstructed direction, and hence $\Psi$ varies from event to event. Thus, we denote the resulting probability density of opening angles by $f (\Psi | E_\nu)$, where we note that $\Psi$ is the random variable and $E_\nu$ is held fixed. For each $E_\nu$ this distribution is normalized such that
\be
\int_0^\pi d\Psi\ f (\Psi | E_\nu) = 1\,.
\ee
We define the $q\%$ containment angle $\psi_q (E_\nu)$ as the opening angle within which $q\%$ of reconstructed events lie at fixed true neutrino energy $E_\nu$. Therefore, $\psi_q$ can also be defined as
\be
\frac{q}{100} = \int_0^{\psi_q (E_\nu)} d\Psi f(\Psi | E_\nu)\,,
\ee
where $q$ is a fixed percentage containment and the integration is over the event-level opening angle $\Psi$ for a fixed $E_\nu$. Hence, for example, $\psi_{50}$ is the opening angle within which $50\%$ of events with true neutrino energy $E_\nu$ are reconstructed. Ref.~\cite{IceCube:2023ame} publicly provides the $q = (20, 50, 80)$ percent containment angles as a function of $E_\nu$.

To construct a continuous opening angle distribution from the three publicly available containment curves ($\psi_{20}$, $\psi_{50}$, and $\psi_{80}$), we approximate $f(\Psi | E_\nu)$ by a log-normal distribution
\be
\label{eq:log_norm_psi}
\! \! \! \!  
f (\Psi | E_\nu)  \!=  \!\frac{1}{\sqrt{2 \pi} \sigma_\Psi (E_\nu) \Psi} \exp \! \left\{ \! -\frac{\big[ \!\ln \Psi - \mu_\Psi(E_\nu) \big]^2}{2 \sigma_\Psi^2 (E_\nu)} \!\right\} \!, \!\! 
\ee
where $\mu_\Psi$ and $\sigma_\Psi$ are the mean and standard deviation of the $\ln (\Psi)$ distribution at a fixed $E_\nu$. For this distribution the containment angle is given by
\be
\label{eq:psi_q}
\psi_q (E_\nu) = \exp \left[ \mu_\Psi(E_\nu) + \sigma_\Psi (E_\nu) \Phi^{-1}(q/100) \right]\,,
\ee
where $\Phi^{-1}$ is the inverse CDF of a standard normal distribution. Furthermore, the median containment angle fixes the mean: 
\be
\label{eq:median}
\mu_\Psi(E_\nu) = \ln \psi_{50}(E_\nu)\,,
\ee
while the separation between the $20\%$ and $80\%$ containment angles determines the logarithmic width $\sigma_\Psi$
\be
\label{eq:sigma}
\sigma_{\Psi} (E_\nu) = \frac{\ln \big[ \psi_{80}(E_\nu)/\psi_{20}(E_\nu) \big]}{ \Phi^{-1}(0.8) - \Phi^{-1}(0.2) }\, ,
\ee
so  Eqs.~\eqref{eq:median} and~\eqref{eq:sigma} fully specify the opening-angle distribution used in our calculation at each $E_\nu$.

The log-normal approximation can be tested directly against the three publicly reported containment curves. Owing to the symmetry of the standard normal distribution, we have $\Phi^{-1}(0.2) = - \Phi^{-1}(0.8)$, and Eq.~\eqref{eq:psi_q} then implies $\psi_{50} = \sqrt{\psi_{20} \psi_{80}}$. We compare this prediction, constructed from the reported $20\%$ and $80\%$ containment angles, with the independently reported $50\%$ containment angle in the \emph{left} panel of Fig.~\ref{fig:ang_recon}. Over $5\ {\rm TeV} \leq E_\nu \leq 100\ {\rm TeV}$, the two agree with a median (maximum) fractional difference of 1.9\% (4.1\%), supporting the use of a log-normal approximation for the opening-angle distribution. Furthermore, Ref.~\cite{IceCube:2026plr} reports an improvement in the shower angular resolution by a factor of approximately $1.5 – 2$ over the energy range relevant to the analysis. We therefore model the updated angular response by rescaling the containment angles of Ref.~\cite{IceCube:2023ame} by a factor of two, $\psi_q(E_\nu)\to\psi_q(E_\nu)/2$. The derived constraints in Fig.~\ref{fig:ann_lim_chisq} are only weakly dependent on this choice and adopting an intermediate improvement factor of $1.75$ changes the limits by $<5\%$, while the most conservative assumption of no improvement weakens them by $\sim 20 – 35\%$ over $m_\chi \simeq 10 – 300$ TeV.

The opening angle distribution specifies the angular distance of the reconstructed direction from the true direction but does not by itself specify its orientation about $\mathbf{n}$. We assume that, at fixed $\Psi$, the reconstructed direction is uniformly distributed in the azimuthal angle $\varphi \in \left[ 0,2\pi \right.)$ about the true direction. Therefore, the probability density per unit reconstructed solid angle is 
\be
\mathcal{R}_{\rm ang} (\widehat{\Omega} \mid \Omega, E_\nu) = \frac{f(\Psi|E_\nu)}{2 \pi \sin \Psi}\,,
\ee
where $\Omega(\ell,b)$ and $\widehat{\Omega}(\widehat{\ell},\widehat{b})$ are the true and reconstructed solid angles respectively. Note that the $1/\sin \Psi$ term converts the 1-D probability density in opening angle to a probability density per unit solid angle, since
\be
d \widehat{\Omega} = \sin \Psi\ d\Psi\ d\varphi\, ,
\ee
and we require
\be
\int d\widehat{\Omega}\ \mathcal{R}_{\rm ang} (\widehat{\Omega} \mid \Omega, E_\nu) = 1\, ,
\ee
so the angular probability function from Eq.~\eqref{eq:pang} is
\be
\epsilon_i^{\rm ang}(E_\nu,\Omega)
=\int_{\widehat{\ell}\in\Delta\widehat{\ell}_i,\,|\widehat{b}|<15^\circ}
d\widehat{\Omega}\ \mathcal{R}_{\rm ang}\left(\widehat{\Omega}\mid\Omega,E_\nu\right),
\label{eq:angular_selection_explicit}
\ee
which describes the probability that  events with true direction $\Omega=(\ell,b)$ and true energy $E_\nu$ will reconstruct in longitude bin $i$ while satisfying the IceCube Galactic-plane selection criteria.
This expression makes explicit that the spatial cuts are imposed on the reconstructed coordinates $(\widehat{\ell},\widehat{b})$, whereas $(\ell,b)$ label the true arrival direction and are held fixed when evaluating the reconstruction probability.

In practice, we evaluate Eq.~\eqref{eq:angular_selection_explicit} using Monte Carlo techniques. For each fixed $E_\nu$ and true direction $\Omega$, we draw an opening angle $\Psi$ from the log-normal distribution in Eq.~\eqref{eq:log_norm_psi} and an azimuth $\varphi$ uniformly from $[0,2\pi)$. These two quantities specify a reconstructed direction $\widehat{\Omega}$ at angular distance $\Psi$ from $\Omega$. The fraction of sampled reconstructed directions satisfying $|\widehat{b}|<15^\circ$ and $\widehat{\ell}\in\Delta\widehat{\ell}_i$ gives $\epsilon_i^{\rm ang}(E_\nu,\Omega)$.

The angular integral in the event prediction of Eq.~\eqref{eq:N_i} is performed over the full true sky, with no restriction on the true Galactic coordinates $\ell$ or $b$. The cuts $|\widehat{b}|<15^\circ$ and $\widehat{\ell}\in\Delta\widehat{\ell}_i$ enter only through the reconstructed-direction probability in Eq.~\eqref{eq:angular_selection_explicit}. The forward folding therefore accounts both for events whose true directions lie inside the Galactic-plane region but reconstruct outside it, and for events originating outside the region that migrate into it; this also accounts for migration between neighboring reconstructed-longitude bins.

We have verified the numerical convergence with respect to both the number of angular Monte Carlo samples and the resolution of the true-sky integration. In particular, refining the true-sky grid from $0.5^\circ$ to $0.25^\circ$ changes the predicted central-bin contribution by less than a few percent. The effect of angular reconstruction on the predicted DM longitude distribution is illustrated in the \emph{right} panel of Fig.~\ref{fig:ang_recon}. Finite angular resolution smooths the intrinsically centrally concentrated DM morphology and redistributes events from the central reconstructed-longitude bin into neighboring bins. In addition, migration across the $|\widehat{b}|=15^\circ$ boundary changes the total number of events satisfying the reconstructed spatial selection relative to the perfect-angular-reconstruction limit.

\begin{figure}
\centering
\includegraphics[width=0.49\textwidth]{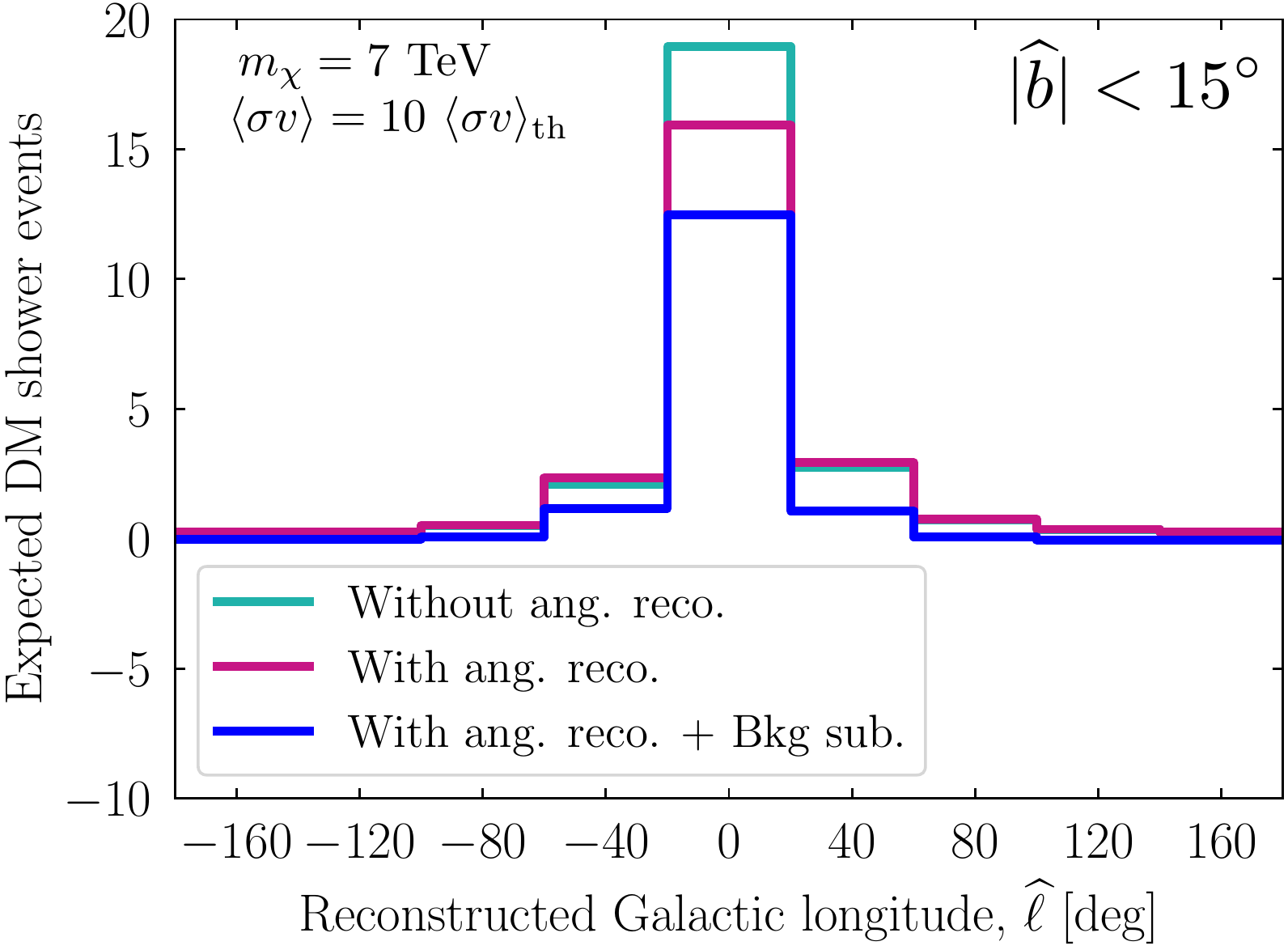}
\caption{\label{fig:bkg_sub_effect}Effect of angular reconstruction and data-driven background subtraction on the predicted DM shower-event distribution in reconstructed Galactic longitude. Angular reconstruction smooths the centrally peaked DM distribution, while the IceCube on–off subtraction further reduces the signal by subtracting the DM contribution present in the off-plane control region.
}
\end{figure}
\begin{figure*}
\centering
\includegraphics[width=0.49\textwidth]{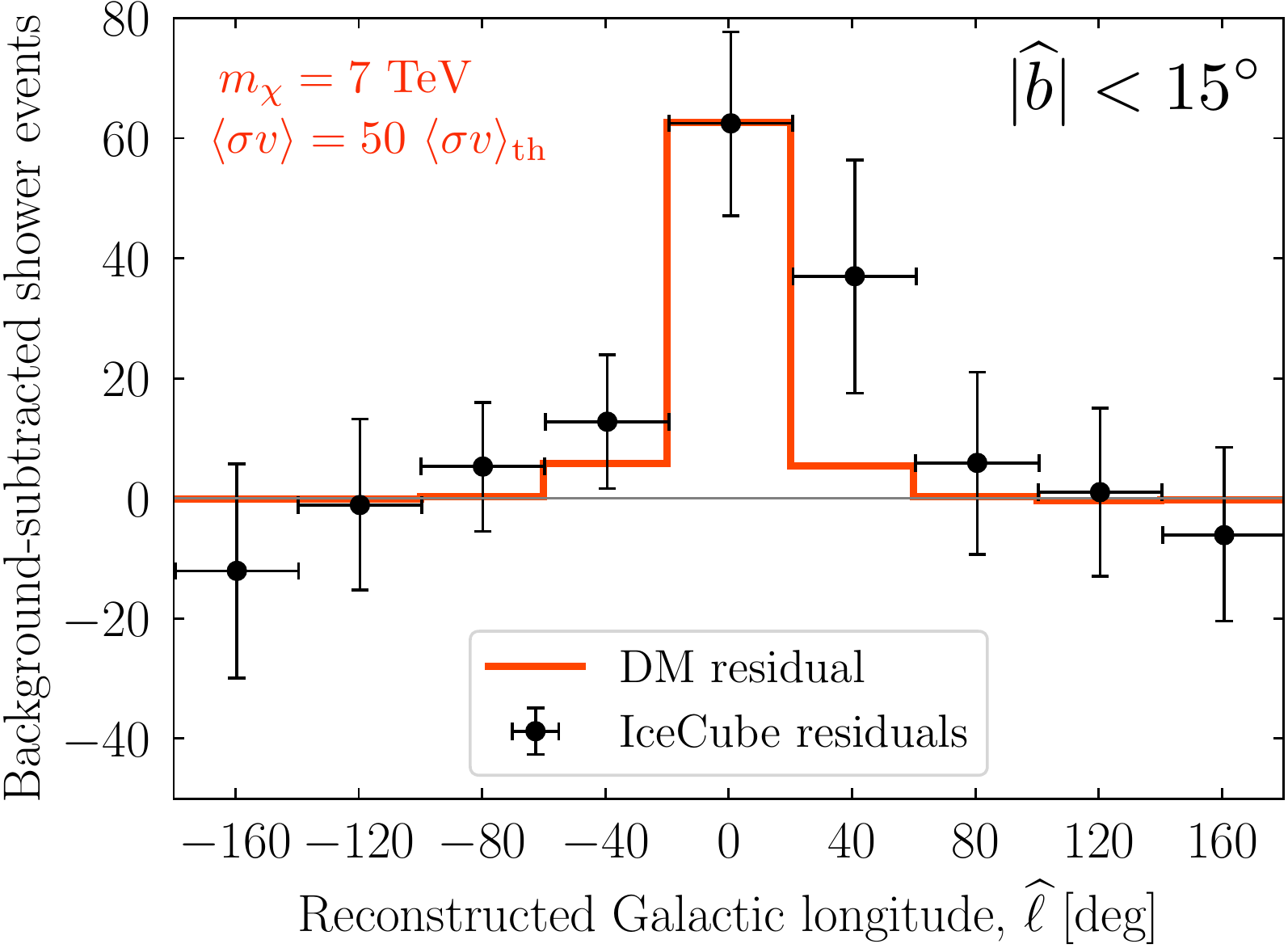}
\includegraphics[width=0.49\textwidth]{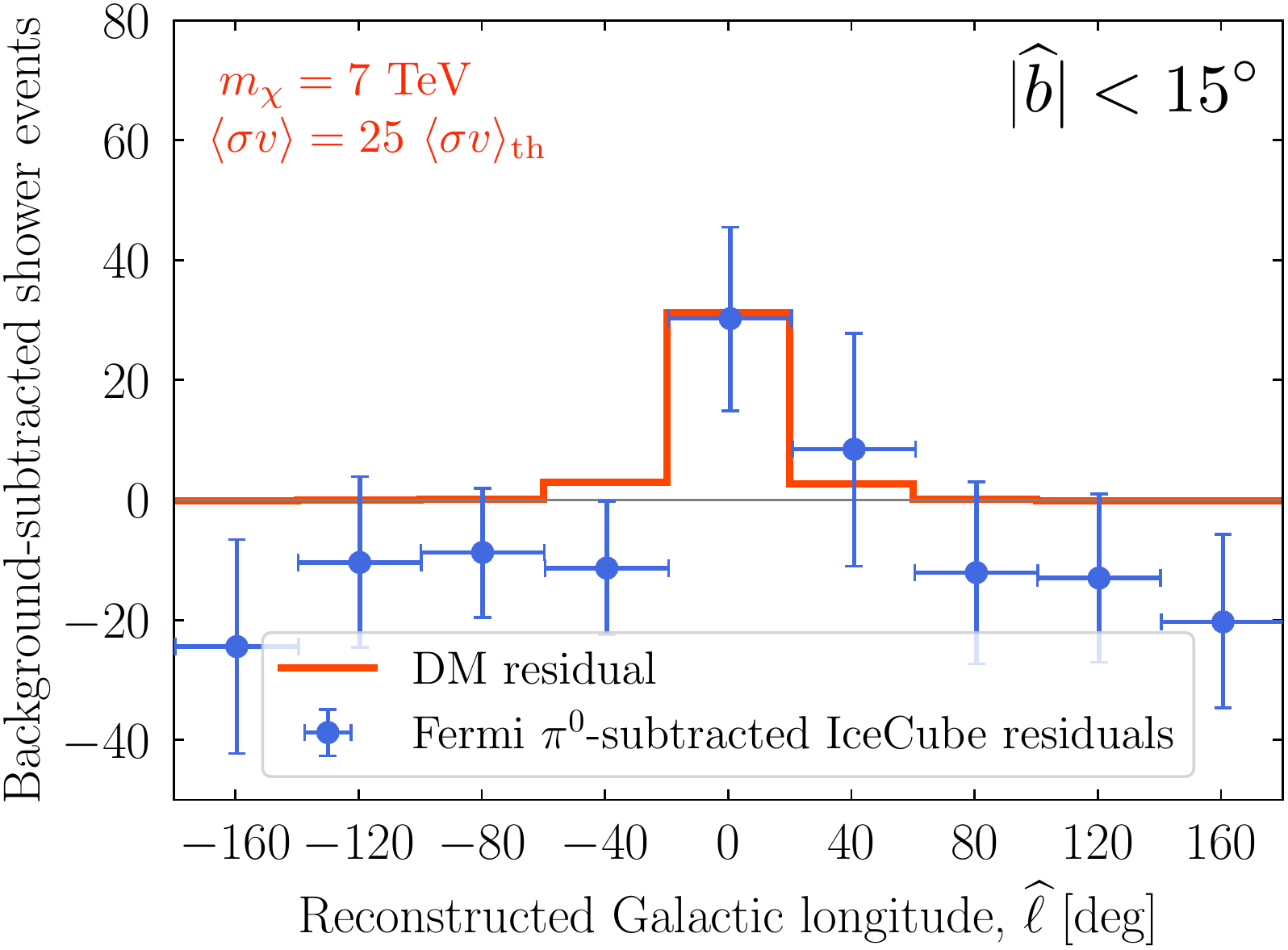}
\caption{\label{fig:residual_vis} Illustration of the spatial analyses used to derive the limits in Fig.~\ref{fig:ann_lim_chisq}. \emph{Left: }Model-independent analysis using the background-subtracted IceCube shower event residual. \emph{Right: }Corresponding analysis after subtraction of the IceCube best-fit Fermi $\pi^0$ Galactic CR template. In both panels, the orange histogram shows the predicted DM residuals after angular reconstruction and data-driven background subtraction (Eq.~\ref{eq:nchi_res}) for $m_\chi = 7\ \rm{TeV}$. The characteristic velocity-averaged annihilation cross section is given by $\langle \sigma v \rangle_{\rm th} \approx 2 \times 10^{-26}\ {\rm cm^3 s^{-1}}$.
}
\end{figure*}

\textit{\textbf{Data-driven background subtraction.}} The Galactic-longitude residuals used in our spatial analysis are obtained by IceCube from reconstructed shower events with $\widehat{E}_\nu > 5\ {\rm TeV}$. In each $40^\circ$ bin of reconstructed Galactic longitude $\widehat\ell$, the number of events reconstructed within the Galactic Plane, $|\widehat{b}| < 15^\circ$, is compared with a data-driven estimate of the background. Rather than assuming a physical model for this background, IceCube estimates it directly from the observed events outside the reconstructed Galactic Plane, $|\widehat{b}| > 15^\circ$, appropriately rescaled to the solid angle of the on-plane region. Since the detector acceptance varies with declination, this estimate is constructed separately in intervals of reconstructed declination with $\Delta \sin \widehat{\delta} = 0.05$, before being summed to obtain the background in each longitude bin.

An important consequence of this procedure is that any DM events present in the off-plane data also contribute to the inferred background. Indeed, the reconstructed off-plane data contain schematically
\be
N_{\rm off}^{\rm data} = N_{\rm off}^{\rm bkg} + N_{\rm off}^{\chi}\,.
\ee
Since IceCube does not distinguish these components when constructing its data-driven background estimate, the contribution $N_{\rm off}^{\chi}$ is also partially subtracted from the on-plane DM signal. Therefore, the DM prediction appropriate for comparison with the published residuals is not simply the number $N_i^{\chi,\rm on}$ of DM events reconstructed on-plane in longitude bin $i$. Instead, we apply the same on-off subtraction to the predicted DM event distribution.

More explicitly, we divide each reconstructed-longitude bin $i$ into the same reconstructed $\sin\widehat\delta$ intervals, labeled by index $a$, used in the IceCube background estimate. In each $(i,a)$ bin, we calculate the predicted numbers of DM events reconstructed in the on-plane and off-plane regions, $N_{ia}^{\chi,\rm on}$ and $N_{ia}^{\chi,\rm off}$, respectively. The off-plane contribution is rescaled to the solid angle of the corresponding on-plane region, giving
\be
\label{eq:nchi_res}
N_i^{\chi,\rm res}
=
\sum_a
\left[
N_{ia}^{\chi,\rm on}
-
\left(\frac{\Delta\Omega_{ia}^{\rm on}}
     {\Delta\Omega_{ia}^{\rm off}}
     \right)
N_{ia}^{\chi,\rm off}
\right] ,
\ee
where $\Delta\Omega_{ia}^{\rm on}$ and $\Delta\Omega_{ia}^{\rm off}$ are the corresponding reconstructed on- and off-plane solid angles. Thus, $N_i^{\chi,\rm res}$ is the part of the DM signal that would remain in the IceCube residual after the same data-driven background subtraction. To evaluate Eq.~\eqref{eq:nchi_res}, we begin with the DM annihilation signal over the full true sky and forward fold it through the shower angular response. The $J$-factor is therefore evaluated at the true coordinates $(\ell,b)$ over the full sky, while the on-plane/off-plane selection and longitude binning are imposed only on the reconstructed coordinates $(\widehat\ell,\widehat b)$. This procedure accounts both for DM events that migrate across the $|\widehat b|=15^\circ$ boundary because of finite angular resolution and for genuine DM emission originating outside the Galactic Plane. We use the resulting $N_i^{\chi,\rm res}$ in the spatial $\chi^2$ analysis.

Figure~\ref{fig:bkg_sub_effect} illustrates the successive effects of angular reconstruction and the data-driven background subtraction on the predicted DM longitude distribution. Finite shower angular resolution smooths the strongly peaked DM morphology, reducing the central-bin event yield and redistributing events into neighboring longitude bins. Applying the IceCube on–off subtraction further reduces the central excess because DM emission is also present in the reconstructed off-plane control region and therefore contributes to the inferred background. At large Galactic longitudes, where the DM surface brightness varies only weakly between the on- and off-plane regions, the two contributions are nearly equal after the solid-angle rescaling and largely cancel. The resulting DM residual therefore approaches zero in the outer longitude bins; small negative values can occur when the rescaled off-plane contribution slightly exceeds the on-plane contribution in a given bin.

\textit{\textbf{Evaluation of the spatial constraints.}} In this section, we provide additional details on the computation of the spatial constraints in Fig.~\ref{fig:ann_lim_chisq}. The reconstructed Galactic longitude distribution of the residual (background-subtracted) shower events as reported by IceCube is shown in Fig.~\ref{fig:residual_vis}. As described above, both the data and the DM prediction are expressed in the same reconstructed spatial observables, that is, the events are binned in reconstructed Galactic longitude $\widehat{\ell}$, the Galactic Plane selection is imposed at $|\widehat{b}| < 15^\circ$, and the DM prediction is subjected to the same data-driven on-off subtraction applied to the IceCube data. The constraints are obtained from the $\chi^2$ procedure described in Eqs.~\eqref{eq:chi2} and~\eqref{eq:chi2-criterion}.

The \emph{left} panel shows the model-independent analysis. The black points denote the background-subtracted IceCube shower event residuals as a function of reconstructed Galactic longitude $\widehat{\ell}$, while the orange histogram shows the corresponding DM residual prediction for $m_\chi = 7\ {\rm TeV}$ and  $\langle \sigma v \rangle = 1 \times 10^{-24}\ {\rm cm^3s^{-1}}$. The underlying DM annihilation intensity is strongly peaked toward the Galactic Center, and after angular reconstruction and the data-driven background subtraction the resulting DM residual remains largest in the central reconstructed-longitude bin, while the contribution at larger $|\widehat{\ell}|$ is strongly suppressed. Consequently, the sensitivity of the model-independent analysis is driven primarily by the central bin, although the $\chi^2$ is evaluated over the full nine-bin reconstructed-longitude distribution. Increasing the annihilation cross section scales the normalization of this DM residual morphology and eventually worsens the agreement with the measured distribution, leading to the $90\%$ CL limit shown in Fig.~\ref{fig:ann_lim_chisq}.

\begin{figure}[t!]
\centering
\hspace{-0.5cm}
\includegraphics[width=0.49\textwidth]{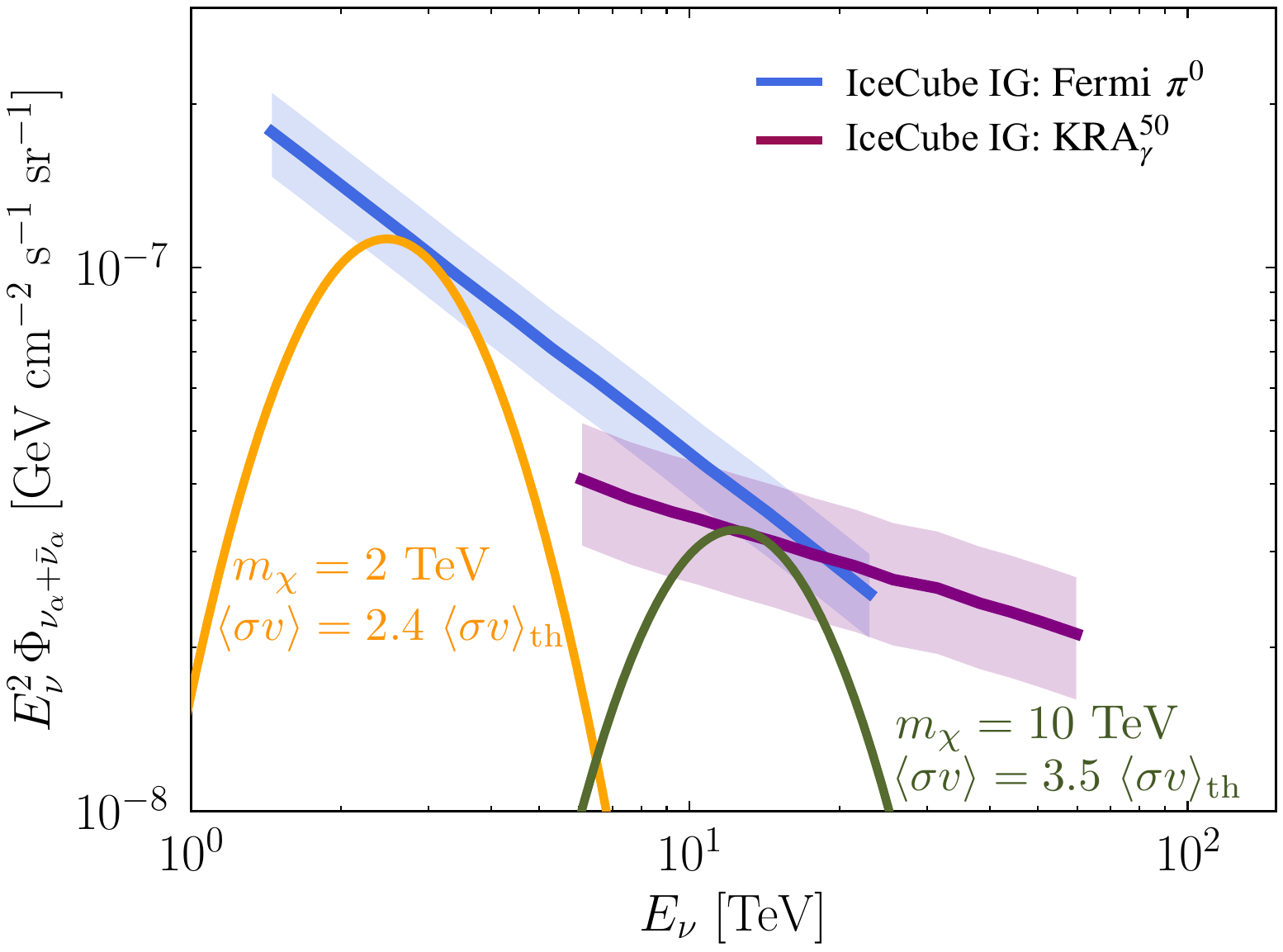}
\caption{\label{fig:envelope_vis} Illustration of the spectral-envelope method used for the comparison shown in Fig.~\ref{fig:ann_lim_envelope}. We show a conservative prescription, where the predicted DM spectrum is compared directly with the IceCube best-fit Inner Galaxy neutrino intensity. The neutrino spectra are shown for representative DM masses and $\langle \sigma v \rangle$. The characteristic velocity-averaged annihilation cross section is given by $\langle \sigma v \rangle_{\rm th} \approx 2 \times 10^{-26}\ {\rm cm^3 s^{-1}}$.
}
\end{figure}

The \emph{right} panel illustrates the corresponding template-subtracted analysis for the Fermi $\pi^0$ Galactic CR emission model. The blue points show the residuals obtained after subtracting the IceCube best-fit Fermi $\pi^0$ contribution from the background-subtracted IceCube shower event residuals in each reconstructed longitude bin. In this analysis, the Fermi $\pi^0$ template normalization is fixed to its IceCube best-fit value. Accordingly, the template normalization is not profiled or varied in this analysis, and the statistical uncertainties are those of the original IceCube residuals. Since a significant fraction of the observed Galactic emission is already accounted for by the astrophysical template, the residual signal available for an additional DM contribution is correspondingly reduced. As a result, the template-subtracted analysis yields stronger limits than the model-independent analysis, while still exploiting the full reconstructed-longitude morphology of the DM residual, as shown in Fig.~\ref{fig:ann_lim_chisq}. For illustration, we keep the same $m_\chi$, and show the DM residual prediction for $\langle \sigma v \rangle = 5 \times 10^{-25}\ {\rm cm^3s^{-1}}$.

The $90\%$ CL limits shown in Fig.~\ref{fig:ann_lim_chisq} are determined from $\chi^2\big(\langle\sigma v\rangle_{\rm lim}\big)-\chi^2_{\rm min}=2.71$, with the $\chi^2$ evaluated over all nine reconstructed-longitude bins. Consequently, the DM residual prediction at the limiting cross section need not coincide with, or lie below, the measured residual in every individual bin. Rather, the limit is determined by the change in the total $\chi^2$ across the full reconstructed longitude distribution.

\vspace{-1.5em}

\section{Additional details on the template-based spectral comparison}
In Fig.~\ref{fig:envelope_vis} we illustrate the spectral-envelope comparison used to define the curves shown in Fig.~\ref{fig:ann_lim_envelope}. We show two representative Galactic CR templates, Fermi $\pi^0$ and KRA$_\gamma^{50}$. The shaded regions denote the $1\sigma$ uncertainty on the corresponding IceCube Inner Galaxy spectra. The neutrino spectra from DM annihilation for $m_\chi = 2\ {\rm TeV}$ and $10\ {\rm TeV}$ are also shown.  This provides a numerically convenient representation of the injected line on a logarithmically spaced energy grid and gives the finite width of the DM spectra shown in Fig.~\ref{fig:envelope_vis}.
For each $m_\chi$, the annihilation cross section is increased until the predicted $\Phi_\nu^\chi$ saturates the spectral envelope according to Eq.~\eqref{eq:flux-comparison}. The different energy ranges probed by the Fermi $\pi^0$ and KRA$_\gamma^{50}$ spectra also illustrate why different templates provide the greatest sensitivity at different DM masses.
\end{document}